\documentclass[runningheads]{llncs}

\usepackage{amsmath,amssymb,amsfonts}
\usepackage{graphicx}
\usepackage{textcomp}                   
\usepackage[table]{xcolor}
\usepackage{booktabs}
\usepackage{array}
\usepackage{listings}
\usepackage{url}
\usepackage[normalem]{ulem}
\usepackage{tikz}
\usetikzlibrary{arrows.meta,positioning,shapes,fit,backgrounds,calc}
\usepackage[hidelinks]{hyperref}
\usepackage{xspace}
\usepackage{microtype}
\usepackage{pgfplots}
\pgfplotsset{compat=1.18}
\usepackage[utf8]{inputenc}
\usepackage[greek,english]{babel} 
\usepackage{xcolor}

\lstdefinelanguage{mfotl}{
  keywords={FORALL, EXISTS, ONCE, EVENTUALLY, ALWAYS, SINCE, UNTIL,
            PREV, NEXT, IMPLIES, AND, OR, NOT, TRUE, FALSE},
  keywordstyle=\bfseries,
  morecomment=[l]{//},
  morestring=[b]",
  sensitive=true
}
\lstdefinelanguage{Python}{
  keywords={def, for, while, if, elif, else, in, return, import, from, as,
            class, with, try, except, raise, pass, lambda, None, True, False,
            and, or, not, is, yield, break, continue, global, nonlocal},
  keywordstyle=\bfseries,
  morecomment=[l]{\#},
  morestring=[b]",
  morestring=[b]',
  sensitive=true
}
\usepackage{float}
\usepackage{bbding}   % provides \Envelope for marking the corresponding author

\usepackage{orcidlink}
\makeatletter
\@ifundefined{orcidID}%
  {\newcommand{\orcidID}[1]{\,\orcidlink{#1}}}%
  {\renewcommand{\orcidID}[1]{\,\orcidlink{#1}}}
\makeatother

\title{Hybrid Hierarchical Runtime Verification for
       Edge-IoT Security: Combining {MonPoly} and {RTLola}\thanks{This is a preprint of an article published in Availability, Reliability and Security (ARES 2026), EU Projects Symposium Workshops, Springer LNCS. The final authenticated version is available online at \url{https://doi.org/10.1007/978-3-032-37218-5_14}.}}
\titlerunning{Hybrid Hierarchical RV for Edge-IoT Security}

\author{%
Nikolaos Kekatos\,\textsuperscript{(\Envelope)}\inst{1}\orcidID{0009-0000-2358-4270} \and
Marinelio Chintri\inst{2}\orcidID{0009-0009-3923-1790} \and
Panagiotis Katsaros\inst{3}\orcidID{0000-0002-4309-5295} \and
Alexios Lekidis\inst{4}\orcidID{0000-0002-8243-7075} \and
Tom Nianios\inst{1}\orcidID{0009-0008-2238-3555} \and
Ioannis Seitoglou\inst{2}\orcidID{0009-0002-5807-3635} \and
Anastasios Temperekidis\inst{1}\orcidID{0009-0005-5856-2780} \and
Stylianos Basagiannis\inst{2}\orcidID{0000-0002-4513-0541}}
\authorrunning{N. Kekatos et al.}
\institute{%
$^{1}$Clone Systems, Larnaca, Cyprus, \email{\{nkekatos,tnianios,atemperekidis\}@clone-systems.com} \quad
$^{2}$International Hellenic University, Serres, Greece, \email{\{basagiannis,marchint,ioaseito\}@ihu.gr} \quad
$^{3}$Aristotle University of Thessaloniki, Greece, \email{katsaros@csd.auth.gr} \quad
$^{4}$University of Thessaly, Larissa, Greece, \email{alekidis@uth.gr}\\[2pt]
\textsuperscript{(\Envelope)}\,Corresponding author.}

\begin{document}
\maketitle

% ============================================================================
\begin{abstract}
Security monitoring of edge-IoT fleets faces three structural
challenges.  (i) A per-node monitor is cheap but cannot see
attacks that coordinate across devices.  (ii) A cloud monitor
sees the full fleet but pays for that view in bandwidth.
(iii) Even at the cloud, a monitor built on a single RV engine
can be fooled by an attacker who compromises a device, raises
one malicious request, and then goes silent: once the events
stop, an event-triggered monitor has nothing left to evaluate.
We propose a three-layer hierarchical runtime-verification
framework that addresses all three.  The edge layer
classifies events as they happen, the gateway layer aggregates
short windows of per-device behaviour, and the cloud layer runs
two complementary RV engines.  MonPoly handles first-order temporal
correlation over the merged alert stream: coordinated overflow (which
genuinely quantifies across devices) plus per-device multi-vector APT,
escalation, and persistent-campaign patterns.  RTLola handles a
time-triggered silent-node property that an event-triggered engine
cannot detect within a bounded delay under fleet silence.  We evaluate the framework
on a 15-actor Docker testbed covering eight attack profiles plus
a silent-bypass scenario.  In the controlled labelled testbed, every
device-attributable incident the framework raises names an
attacker-labelled device, and the RTLola
tier catches silent-bypass attempts the event-triggered tier misses.  Per-event monitoring stays in the
microsecond range at the edge and gateway, with low end-to-end
alert-to-incident latency at the cloud.
\end{abstract}

% ============================================================================
\section{Introduction}
\label{sec:intro}

Security-critical decisions in edge-IoT deployments are pushed
ever closer to the device~\cite{cavalli2024iotsec}, but doing
everything at the edge has a structural cost in visibility.  A
per-node monitor sees only its own device, so an attack that
coordinates across devices~\cite{ukraine2015} is invisible to it.
Pushing every
event to a central cloud monitor restores the fleet-wide view,
but pays for it in bandwidth and detection
latency.  And even a well-resourced cloud
monitor built on a single RV engine keeps one blind spot:
silence.  An attacker who compromises the on-device monitor,
issues a single malicious request, and then goes quiet leaves an
event-triggered monitor~\cite{bonakdarpour2013ttrv} with nothing left to evaluate.

Each of these choices fixes one weakness only to expose another, so
neither a single tier nor a single engine suffices.  To cover all
three at once, this paper proposes a three-layer hierarchical
runtime-verification framework built on formal RV
methods~\cite{leucker2009brief,falcone2021taxonomy} that assigns
each tier the checks it can afford and reserves for the cloud the
checks no lower tier can express.  The edge
layer (L1) is a per-event canonicaliser that classifies buffer
overflows, time spoofing, fuzzing, and safe transactions.  The
gateway layer (L2) aggregates short windows per device, covering
repeated overflows, spam and pulsing DoS, and immediate fuzz or
time-anomaly alerts.  The cloud layer (L3) is where the paper makes its
central design choice: rather than pick one RV engine, we run two side by
side.  The first RV engine is MonPoly~\cite{monpoly2014};
its expressive first-order temporal specification language handles four
past-time correlation properties over the merged alert stream.  One of
them, coordinated overflow, genuinely quantifies across devices and is
provably inexpressible at any single-device tier (Theorem~1); the other
three (multi-vector APT, escalation, persistent campaign) are per-device
patterns ($\forall d$) that we consolidate at the cloud by choice.  The second is
RTLola~\cite{rtlola2019}, which handles a single time-triggered
property, silent-node anomaly, whose verdict must be emitted at a fixed
deadline rather than on input.  When fed the merged multi-device stream, an
event-triggered engine like MonPoly stalls only under \emph{total} fleet
silence -- exactly the silent-bypass case -- which RTLola's time-triggered
semantics bounds natively.  The result is that fast, deterministic decisions stay at the edge; per-device temporal
aggregation stays at the gateway; cross-device correlation and
absence-of-event detection live at the cloud.

We evaluate the framework on a 15-actor Docker testbed spanning
eight attack profiles (normal, overflow, time-spoof, spam/DoS,
low-and-slow overflow, fuzz, pulsing DoS, mixed APT).  Single-node
attacks are caught by the edge and gateway tiers; multi-vector
APT, persistent campaigns, and silent-bypass require the hybrid
cloud tier.

\emph{Why three layers.}  Edge-only stream-based RV~\cite{rtlola2019,tessla2018,pike2017copilot}
offers rich per-trace semantics but cannot correlate behaviour
across $N$ devices.  A gateway tier that counts events over short
windows brings repeated overflow and spam into reach.  But it
still misses attacks that mix \emph{different} vectors across
\emph{different} devices in one window.  That is exactly the
pattern multi-vector APT and coordinated campaigns follow, as in
ICS incidents like the 2015 Ukrainian power-grid
attack~\cite{ukraine2015}.
MonPoly's specification language~\cite{monpoly2014} supplies the
quantifiers this class of property needs, but running it on every
raw packet from $N$ devices is too costly in bandwidth and
compute.  That is why correlation sits at the cloud tier.

Hierarchical CPS-RV
literature~\cite{strel2017,sastl2020} establishes the
decomposition pattern; we refine it on four axes.  First,
we prove a layer non-collapse theorem (Theorem~1,
Section~\ref{sec:spec:compose}): under the single-device gateway
abstraction we adopt, no per-device rule can decide a property that
quantifies over two distinct devices, so detecting a coordinated
cross-device attack is provably impossible below the cloud under that
abstraction.  The
cloud tier therefore fills a gap the lower tiers cannot, by
construction, fill.

Second, we identify a gap less often discussed:
event-triggered monitors like MonPoly report a missing event only
once a later-timestamped event arrives, with no bound on detection
latency.  Time-triggered stream monitors like RTLola, by contrast,
emit a verdict at the deadline regardless of input but express
cross-device joins less naturally.  Running both engines side by side covers both
gaps without pushing either outside the properties it handles well.
Third, we contribute an executable artefact that turns the
per-tier specifications into an end-to-end audit trail, tracing every
cloud incident back to the gateway alerts and edge events that
produced it.  Fourth, we report measured per-event monitoring
cost at each tier and end-to-end alert-to-incident latency at the
cloud (Section~\ref{sec:eval:cost}), making these deployment choices
quantifiable.

\emph{Cost asymmetry and adversary model.}  The framework forces an attacker to pay coordination cost across
at least three devices, while the defender pays only a bounded
per-event canonicaliser at the edge (mean ${\sim}57\,\mu$s) and a
sliding-window update at the gateway (mean ${\sim}710\,\mu$s,
Section~\ref{sec:eval:cost}), with constant-size witness sets per
alert.  On our workload the gateway-to-cloud alert stream is
roughly $7\times$ smaller than the edge input; we do not claim a larger
reduction, as that would assume a raw-packet upstream our testbed
does not measure.  End-to-end cloud latency stays around 1.7\,s,
with the silent-node detector adding a fixed five-second window.

We assume an adversary who compromises one or more edge
monitors and knows the framework's thresholds and the cloud
property catalogue.  ``Success'' for the adversary means causing
exfiltration or disruption without raising any cloud-level MonPoly
or RTLola incident.  Supply-chain compromise of the
edge or gateway binaries (assumed handled by attestation),
gateway-timestamp tampering, and direct interference with the
edge sidecar are out of scope.

\emph{Contributions.}
The paper makes four contributions.  We give a three-layer
formal specification with thirteen properties (P1.1--P1.4 at the
edge, P2.1--P2.4 at the gateway, and P3.1--P3.5 at the cloud)
that together cover eight attack classes plus a silent-bypass
scenario.  We describe a hybrid reference architecture in which
the cloud tier runs four MonPoly processes alongside one
RTLola time-triggered process.  We instantiate the architecture
on a 15-actor Docker testbed, and we evaluate it by reporting
per-property incident counts, per-tier monitoring cost,
alert-to-incident latency, and a replay of real IoT-23 data.

% ============================================================================
\section{Attack-Type Catalogue}
\label{sec:attacks}

The framework is evaluated on eight IoT attack profiles drawn
from the attack-surface taxonomy in edge-IoT-security
surveys~\cite{cavalli2024iotsec}.  Each profile is realised as a
Python device emulator with controlled parameters
(Table~\ref{tab:attacks}).  The eight profiles are deployed
across 15 containers via \texttt{docker-compose}: five clean
baselines (\textsc{node-1}--\textsc{node-5}) and ten attackers
(\textsc{node-6}--\textsc{node-15}), one of each non-clean
profile plus a mix of mixed-APT actors; ground-truth labels are
recorded at the generator.

\begin{table}[t]
\centering
\caption{The eight attack profiles emulated by the testbed (A1--A8) plus the silent-bypass scenario (AS1), with the detecting tier/property. AS1 is not a separate node: it arises as a behaviour of existing attackers.}
\label{tab:attacks}
\setlength{\tabcolsep}{4pt}
\renewcommand{\arraystretch}{1.15}
\footnotesize
\begin{tabular}{@{}l l l l@{}}
\toprule
\textbf{ID} & \textbf{Profile} & \textbf{Parameters} & \textbf{Detected by} \\
\midrule
A1 & Normal traffic         & $[10,30]$\,B, interval $U(0.1,4.0)$\,s  & (baseline) \\
A2 & Buffer overflow        & $[35,60]$\,B; 80\% \textsf{vulnerable\_send}           & L1 / P1.1 \\
A3 & Time spoof             & \textsf{turn} offset $\pm 300{/}600$\,s               & L1 / P1.2; L2 / P2.3 \\
A4 & Spam DoS               & interval $U(0.005,0.05)$\,s, \textsf{safe\_send}       & L2 / P2.2 \\
A5 & Low-and-slow ovfl      & 31--32\,B payloads, 10\% rate                          & L1 / P1.1; L2 / P2.1 \\
A6 & Fuzz                   & malformed content$^{*}$                                & L1 / P1.3; L2 / P2.4 \\
A7 & Pulsing DoS            & 20-event bursts every 20--30\,s                        & L2 / P2.2 \\
A8 & Mixed APT              & 40\% mix of overflow / spoof                           & L3 / P3.1, P3.2 \\
AS1 & Silent bypass       & overflow then silence ($\geq 5$\,s)                   & L3 / P3.5 \\

\bottomrule
\end{tabular}
\\[2pt]
{\footnotesize $^{*}$\,Fuzz payloads are non-printable or structurally invalid.}
\end{table}

The attack profiles partition cleanly by the kind of
quantification needed to detect them: single-event classes
(A2, A3, A6) by an L1 per-event predicate; time-windowed
classes (A4, A5, A7) by L2 sliding counts; the multi-vector class (A8) by L3's parametric quantification
-- a logical necessity for the cross-device form P3.2 under the single-device gateway abstraction (Theorem~1), and an engineering choice for the per-device properties P3.1, P3.3, P3.4 -- and the silent-bypass scenario (AS1) by the time-triggered P3.5.  Empirical distributions appear in
Tables~\ref{tab:aggregate} and~\ref{tab:l3-detection}.
% ============================================================================
\section{Three-Layer Specification}
\label{sec:spec}

The framework is given as a triple of formal specifications
$\Phi = \langle \varphi_{L1}, \varphi_{L2}, \varphi_{L3} \rangle$,
each scoped to its tier's data granularity: L1 is per-event,
L2 is per-window-per-device, L3 is multi-device past-time.  We
write all specifications in standard temporal logic for
engine-independence; the reference implementation
(Section~\ref{sec:arch}) realises them as Python stream
processors at L1 and L2 and as MonPoly~\cite{monpoly2014}
metric first-order temporal logic (MFOTL) formulas at L3.

\subsection{Event schema}
\label{sec:spec:schema}
Four event types flow through the pipeline.  A raw device event
$e_i = \langle\textsf{turn},\textsf{actor},\textsf{tool},
\textsf{args}\rangle$ has a \textsf{tool} field over \{safe,
vulnerable, time-spoof, fuzz\} sends and an \textsf{args} record
of attempted-size, actual-sent, buffer-limit and content.  L1 emits one
canonical event per input event (buffer-overflow, time-anomaly,
logic-fuzz, or safe-tx).  L2 emits one of three alert types
(repeated-overflow, dos, immediate) and forwards safe-tx to L3.
L3 emits one of five incidents (apt-indicator,
coordinated-attack, escalation-pattern, persistent-threat, silent-node-anomaly) with
a severity label.  We write $\Sigma_{\mathit{raw}}$,
$\Sigma_{\mathit{canon}}$, and $\Sigma_{\mathit{alert}}$ for the
sets of raw events, canonical events, and alerts, respectively.
For example, a raw \textsf{vulnerable\_send} whose 48\,B payload
exceeds the 30\,B buffer limit ($\Sigma_{\mathit{raw}}$) becomes the canonical event
$\textsf{buffer\_overflow}(\textsf{node-6},48)$
($\Sigma_{\mathit{canon}}$), and three such within 30\,s raise the
alert $\textsf{overflow}(\textsf{node-6})$
($\Sigma_{\mathit{alert}}$).

\subsection{Per-layer property specifications}
\label{sec:spec:layers}
\paragraph{L1: per-event properties (canonicaliser).}
The L1 monitor enforces four memoryless boolean predicates over
the current raw event $e$ (no inter-event state), each emitting
a single canonical event when its body holds:

\begin{itemize}\setlength{\itemsep}{1pt}
\item \textbf{P1.1} (buffer overflow):
$e.\mathsf{tool} = \textsf{vulnerable\_send} \wedge
e.\mathsf{actual\_sent} > e.\mathsf{buffer\_limit}$.
\item \textbf{P1.2} (time spoofing):
$|\mathit{current\_time} - e.\mathsf{turn}| > \tau_{ts}$
($\tau_{ts}{=}15$\,s).
\item \textbf{P1.3} (input fuzzing):
$e.\mathsf{tool} = \mathsf{fuzz\_send} \wedge \neg\,\mathsf{wellformed}(e.\mathsf{content})$,
where $\mathsf{wellformed}(c)$ holds iff $c$ conforms to the expected payload
contract (a printable-ASCII string within the valid length domain). Oversized payloads ($e.\textsf{actual\_sent} > e.\textsf{buffer\_limit}$) are by construction the buffer-overflow class (P1.1); fuzzing here is value/content well-formedness, so the fuzz and overflow domains are disjoint.
\item \textbf{P1.4} (\textsf{safe\_tx}, residual):
$\neg\mathrm{P1.1} \wedge \neg\mathrm{P1.2} \wedge \neg\mathrm{P1.3}$.
\end{itemize}
\noindent
The four predicates are mutually exclusive and exhaustive: every
raw event emits exactly one canonical event, with arguments
$(e.\mathsf{actor},\, x)$ where $x$ is either the violating
\textsf{actual\_sent} (P1.1, P1.3, P1.4) or the suspect
\textsf{turn} (P1.2).  \emph{Semantic note.}  P1.4 is a
\emph{residual} predicate (the event did not match P1.1--P1.3),
not a positive ``verified-safe'' assertion.  Downstream
properties built on \textsf{safe\_tx} (notably P3.3 escalation)
therefore detect transitions from \emph{not-one-of-these-three-anomalies}
to overflow plus time-spoof; they are signature-style temporal
patterns, not general threat-progression detectors.

\paragraph{L2: time-bounded temporal properties (gateway monitor).}
\label{sec:spec:l2}
L2 maintains short-window per-device counters and evaluates
four temporal predicates over sliding windows
($W_{\mathit{ovf}}{=}30$\,s for repeated overflow,
$W_{\mathit{spam}}{=}5$\,s for spam/pulsing, stateless for the
immediate alerts).  Let $\#_W(\pi, d)$ denote the
count of events $\pi(d,\cdot)$ in the past $W$ seconds for device
$d$.

\begin{itemize}\setlength{\itemsep}{1pt}
\item \textbf{P2.1} (repeated buffer overflow):
   \[\square\bigl(\#_{[30s]}(\textsf{buffer\_overflow},d) \geq
     N_{\mathit{ovf}}
     \to \textsf{overflow}(d)\bigr),\]
   with $N_{\mathit{ovf}}=3$ over a 30\,s window; this is the
   gateway rule that catches sustained-overflow attack patterns,
   including the low-and-slow overflow profile (A5).
\item \textbf{P2.2} (spam / pulsing DoS):
   \[\square\bigl(\#_{[5s]}(\textsf{safe\_tx},d) \geq
     N_{\mathit{spam}}
     \to \textsf{dos\_attack\_alert}(d)\bigr),\]
   with $N_{\mathit{spam}}=30$ for sustained spam and a tighter
   $N_{\mathit{pulse}}=15$ over $W_{\mathit{pulse}}=2$\,s for
   pulsing bursts.
\item \textbf{P2.3} / \textbf{P2.4} (immediate alerts):
   any \textsf{time\_anomaly} or \textsf{fuzz} event triggers an
   immediate alert.  These are deliberately stateless: a
   single event suffices because neither has a plausible
   benign-traffic baseline.
\end{itemize}

\paragraph{L3: cross-device correlation properties (MFOTL).}
\label{sec:spec:l3}
L3 consumes the L2 alert stream and evaluates \emph{five}
properties: four event-triggered MFOTL formulas
(P3.1--P3.4) executed by MonPoly and one time-triggered stream
property (P3.5) executed by RTLola.  Let
$\textsf{evt}_i \in \{\textsf{overflow}, \textsf{timespoof},
\textsf{fuzzing}\}$ range over the L2 alert predicates.
\begin{align*}
\textbf{P3.1}: \;& \forall d.\,\square\bigl(\,
   \exists t_1{<}t_2,\, t_2{-}t_1{\leq}60,\,
   \textsf{evt}_1(d,t_1) \wedge \textsf{evt}_2(d,t_2), \\
 & \qquad \textsf{evt}_1{\neq}\textsf{evt}_2
   \to \textsf{apt\_indicator}(d)\bigr) \\
\textbf{P3.2}: \;& \square\bigl(\,\diamond_{[0, 30\mathrm{s}]}\textsf{overflow}(d) \wedge 
   \mathit{cnt} = \#\bigl\{d' \,\big|\,
     \diamond_{[0,30\mathrm{s}]}\textsf{overflow}(d')\bigr\}
   \wedge \mathit{cnt}\geq 3 \\
 & \qquad \to \textsf{coordinated\_attack}(\mathit{d}, \mathit{cnt})\bigr) \\
\textbf{P3.3}: \;& \forall d.\,\square\bigl(\,
   \diamond_{[10,60\mathrm{s}]}\textsf{safe\_tx}(d) \wedge
   \diamond_{[0,50\mathrm{s}]}\textsf{overflow}(d) \wedge \\
 & \qquad \diamond_{[0,50\mathrm{s}]}\textsf{timespoof}(d)
   \to \textsf{escalation\_pattern}(d)\bigr) \\
\textbf{P3.4}: \;& \forall d.\,\square\bigl(\,
   \#_{[1\mathrm{h}]}(\textsf{overflow},d) \geq 5
   \to \textsf{persistent\_threat}(d)\bigr) \\
\textbf{P3.5}: \;& \forall d.\,\square\bigl(\,
   \textsf{overflow}(d,t_0) \wedge
   \neg\diamond_{[t_0, t_0+5\mathrm{s}]}\textsf{safe\_tx}(d) \\
 & \qquad \to \textsf{silent\_node\_anomaly}(d)\bigr)
\end{align*}

\noindent P3.1--P3.4 are MFOTL formulas executed by MonPoly;
P3.5 is the silent-node anomaly. MFOTL can express it as a bounded-response formula and MonPoly supports bounded-future operators; however, an event-triggered monitor advances its clock only with the timestamps of incoming events.  Because MonPoly is fed the merged multi-device stream, any device's alert advances its clock, so its detection latency is unbounded only under \emph{total} fleet silence -- precisely the silent-bypass case.  Injecting synthetic clock-ticks into MonPoly would bound the delay, but only via a timer outside the specification driven by an external scheduler.
RTLola's time-triggered semantics instead emits the verdict at the 5\,s
window closure without input, bounding detection at the window length natively.  Table~\ref{tab:properties}
summarises the thirteen properties.

\begin{table}[t]
\centering
\caption{The thirteen formal properties across the three
monitoring tiers.  $\square$ denotes always; $\diamond_{[a,b]}$
a bounded-past existential and future-bounded for P3.5; $\#_W$ a sliding count.  P3.1--P3.4
use MonPoly (event-triggered); P3.5 uses RTLola (time-triggered).}
\label{tab:properties}
\setlength{\tabcolsep}{4pt}
\renewcommand{\arraystretch}{1.1}
\footnotesize
\begin{tabular}{@{}lllll@{}}
\toprule
\textbf{ID} & \textbf{Layer} & \textbf{Engine} & \textbf{Type} & \textbf{Targets} \\
\midrule
P1.1  & L1 & Python    & Per-event       & buffer overflow \\
P1.2  & L1 & Python    & Per-event       & time spoofing  \\
P1.3  & L1 & Python    & Per-event       & fuzzing \\
P1.4  & L1 & Python    & Per-event       & safe transactions \\
\midrule
P2.1  & L2 & Python    & Temporal (30\,s) & repeated overflow \\
P2.2  & L2 & Python    & Temporal (5\,s)  & spam / pulsing DoS \\
P2.3  & L2 & Python    & Immediate       & time anomaly \\
P2.4  & L2 & Python    & Immediate       & fuzz events \\
\midrule
P3.1  & L3 & MonPoly   & MFOTL (60\,s)   & multi-vector APT \\
P3.2  & L3 & MonPoly   & MFOTL count \& witness (30\,s) & coordinated attack \\
P3.3  & L3 & MonPoly   & MFOTL (10--60\,s) & escalation pattern \\
P3.4  & L3 & MonPoly   & MFOTL count (1\,h) & persistent campaign \\
P3.5  & L3 & RTLola    & Time-triggered (5\,s) & silent-node anomaly \\
\bottomrule
\end{tabular}
\end{table}

\subsection{Hierarchical composition}
\label{sec:spec:compose}
The three layers form a strict containment hierarchy by data
locality: L1 sees one event at a time; L2 sees one device's
events over a few seconds; L3 sees alert summaries across all
devices over minutes-to-hours.  Composition is operationally a
streaming pipeline: L1 emits canonical events forwarded to L2;
L2 emits alert events forwarded to L3; L3 emits incident events
to the audit log.  Each layer attaches a witness sub-trace
(layer, device, predicate, timestamp) to every output, so an L3
incident is mechanically traceable back through the contributing
L2 alerts and the contributing L1 canonical events.
In particular, P3.2 incidents carry the full contributing device set as a runtime witness emitted directly from MonPoly's satisfying assignments.
  L1 alone
classifies the four per-event categories (three single-event attack
classes plus the residual \textsf{safe\_tx}); L2 adds temporal patterns
invisible at the per-event level; L3 adds the cross-device
correlations that no single-device monitor can express, a
claim Theorem~1 below makes precise.  The split is
asymmetric in deployment cost: per-event work stays on the edge,
per-device temporal aggregation at the gateway, parametric MFOTL
only at the cloud correlator.

\noindent\emph{Layer abstractions.}  Let an L1 monitor be a
total memoryless map (function) $\mu_1:\Sigma_{\mathit{raw}}\to
\Sigma_{\mathit{canon}}$ (P1.1--P1.4 are exhaustive, so every raw
event yields exactly one canonical event), and an L2
\emph{single-device} monitor a map
$\mu_2:(\Sigma_{\mathit{canon}})^{*}_d\to
2^{\Sigma_{\mathit{alert}}}$, where $(\Sigma_{\mathit{canon}})^{*}_d$
is the set of finite canonical-event sequences of a single fixed
device $d$ (with finite sliding-window counters) and
$2^{\Sigma_{\mathit{alert}}}$ the sets of alerts; $\mu_2$'s domain
contains no device variable.  Let
$\mathcal{F}_{\mathit{single}}$ denote the class decidable by any
such $\mu_2$.

\noindent\textbf{Theorem 1 (P3.2 requires cross-device
quantification).}\;
\emph{Property P3.2 quantifies existentially over distinct
devices; even its two-device core ($\exists d_1{\neq}d_2.\, \textsf{overflow}(d_1)
\wedge \diamond_{[0,30s]}\textsf{overflow}(d_2)$) is
not expressible in $\mathcal{F}_{\mathit{single}}$.}
\emph{Proof.}  Suppose for contradiction that
$\psi\in\mathcal{F}_{\mathit{single}}$ decides P3.2.  Consider a
two-device trace $T=T_{d_1}\sqcup T_{d_2}$ where $T_{d_1}$ has
$\textsf{overflow}(d_1)$ at $t_1$ and $T_{d_2}$ has
$\textsf{overflow}(d_2)$ at $t_2$, $|t_2{-}t_1|\leq 30$\,s.  By
construction of $\mathcal{F}_{\mathit{single}}$, $\psi(T)$ is
determined by $\psi(T_{d_1})$ and $\psi(T_{d_2})$ separately:
neither call sees the other device.  Each single-device
projection lacks the second-device witness P3.2 needs, so both
return \textsf{false}, contradicting the assumption.\hfill$\Box$

\noindent Note that the abstraction restricts the L2 \emph{rule class}, not the gateway's physical visibility: the gateway observes all devices, but each rule's state and verdict involve exactly one device, matching the per-device deques of the implementation.

\noindent\emph{Scope.}  A hypothetical multi-device gateway
$\mu_2^{\mathit{multi}}$ that saw all devices' streams could
express P3.2 (with $O(N)$ per-gateway state vs.\ $O(1)$-per-device
for $\mu_2$); P3.1, P3.3, P3.4 lie also outside the
\emph{chosen} L2 fragment (5--30\,s windows, no joint vector
predicates, no hour-scale counts), but not outside every conceivable L2-shaped monitor: their placement at L3 is an engineering choice,
not a logical necessity.
The layering claim is therefore threefold: (i) formal necessity of
cross-device quantification for P3.2 under
$\mathcal{F}_{\mathit{single}}$;  (ii) the choice of the single-device abstraction itself, motivated by
O(1)-per-device gateway state
 and (iii) measured cost
separation (Section~\ref{sec:eval:cost}).

% ============================================================================
\section{Runtime Architecture}
\label{sec:arch}

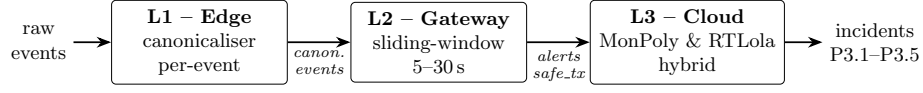
\begin{figure}[t]
\centering
\resizebox{\columnwidth}{!}{%
\begin{tikzpicture}[
  font=\footnotesize,
  tier/.style={draw, rounded corners=2pt, align=center, inner sep=4pt,
               minimum height=11mm, minimum width=28mm,
               font=\footnotesize},
  arrow/.style={-Stealth, thick}
]
\node[tier]    (l1) {\textbf{L1 -- Edge}\\canonicaliser\\per-event};
\node[tier, right=10mm of l1] (l2) {\textbf{L2 -- Gateway}\\sliding-window\\5--30\,s};
\node[tier, right=10mm of l2] (l3) {\textbf{L3 -- Cloud}\\MonPoly \& RTLola\\hybrid};
\node[left=6mm of l1, align=center, font=\footnotesize]  (in)  {raw\\events};
\node[right=6mm of l3, align=center, font=\footnotesize] (out) {incidents\\P3.1--P3.5};
\draw[arrow] (in)  -- (l1);
\draw[arrow] (l1)  -- node[below, align=center, font=\scriptsize\itshape] {canon.\\events} (l2);
\draw[arrow] (l2)  -- node[below, align=center, font=\scriptsize\itshape] {alerts\\safe\_tx} (l3);
\draw[arrow] (l3)  -- (out);
\end{tikzpicture}%
}
\caption{Three-layer runtime architecture.  L1 (edge) classifies
each raw event into a canonical event in constant time.  L2
(gateway, 5--30\,s windows) aggregates canonical events into
repeated-overflow or DoS alerts and forwards both alerts and
safe-transaction events to L3.  L3 (cloud) hosts four persistent MonPoly MFOTL monitors and one RTLola time-triggered monitor that consume the L2 stream incrementally and emit multi-device security incidents (P3.1--P3.5) tagged with
severity.  All three layers run as cooperating processes inside
a \emph{monitor} container hosting L1 and L2 as a Unix pipe, and a separate \emph{correlator} container hosting L3 in the reference implementation; the
boundaries map directly to a physical edge / gateway / cloud
deployment.}
\label{fig:arch}
\end{figure}

We instantiate Section~\ref{sec:spec} in a Docker testbed: one process per
IoT device, a \emph{monitor} container hosting the L1$\to$L2 pipeline as a
Unix pipe between Python stream processors, and a separate \emph{correlator} container hosting the L3 MonPoly and RTLola subprocesses (Figure~\ref{fig:arch}).
 The co-located reference layout maps directly to a physical
deployment, with L1 on each edge device, L2 on a local gateway, L3
in a regional cloud / SOC.  Inter-tier traffic on the measured
workload is dominated by forwarded \textsf{safe\_tx} events
required by P3.3 and P3.5; only the security-alert subset
(941 of 6861 L2 outputs) is
genuinely smaller than the L1 input.

\paragraph{L1 -- edge canonicaliser.}
The L1 monitor is a short Python stream processor that reads
JSON-encoded raw events and emits one canonical event per input
event as a MonPoly-friendly record \texttt{@T predicate("device",
arg)}.  The four predicates P1.1--P1.4 are evaluated in priority
order.

\paragraph{L2 -- gateway temporal monitor.}
The L2 monitor maintains three per-device sliding-window deques
(spam, pulsing, overflow) and evaluates P2.1--P2.4 incrementally;
P2.2 is structurally identical to P2.1 with a \textsf{safe\_tx}
queue and a higher threshold, while P2.3 / P2.4 fire on the first
matching event.  Every L2 output is JSON-serialised to L3.

\paragraph{L3 -- hybrid MonPoly \& RTLola correlator.}
L3 hosts five persistent subprocesses: four MonPoly MFOTL
processes (P3.1--P3.4) fed the full L2 alert stream, and one
online RTLola process hosting P3.5, fed only the \textsf{safe\_tx}
and \textsf{overflow} predicates it references.  Reader threads
parse each subprocess's output, deduplicate by
$(\textit{prop},\textit{device},\lfloor t/W \rfloor)$, and append
to an incident log tagged with the producing engine and severity
({\sc critical} for P3.1/P3.2/P3.5, {\sc high} otherwise).

\paragraph{Orchestrator generalisation.}
The orchestrator (per-engine event filtering, format translation,
subprocess management, verdict aggregation) generalises to
additional RV engines: a third engine needs only its predicate
subset, input encoding, and output regex.  

% ============================================================================
\section{Evaluation}
\label{sec:eval}

We describe the methodology in sufficient detail to support independent
reproduction on the 15-actor Docker testbed packaged in the
accompanying artefact.
All experiments were run on a single MacBook Pro with an Apple~M1 SoC
(8-core CPU) and 8\,GB unified memory. The stack executed under Docker Desktop~29.4.3, whose Linux VM was allocated 8 CPUs and ${\sim}3.8$\,GB RAM.

\subsection{Dataset}
\label{sec:eval:data}

All traces are generated on the Docker-based testbed: fifteen
device emulators (one per actor, configured via per-device
identity and profile environment variables), one \emph{monitor}
container hosting the L1$\to$L2 pipeline, and one \emph{correlator}
container hosting the L3 MonPoly and RTLola subprocesses.  Each node
appends JSON-line \textsf{tool\_call} events with fields
\textsf{turn} (device-reported Unix seconds, spoofable),
\textsf{actor}, \textsf{tool} $\in$
\{\textsf{safe\_send}, \textsf{vulnerable\_send},
\textsf{time\_spoof\_send}, \textsf{fuzz\_send}\}, and
\textsf{args} $=$ $\langle\textsf{attempted\_size},
\textsf{actual\_sent}, \textsf{buffer\_limit}{=}30\rangle$.
The 17 containers share a host-bind-mounted volume for the log
files; logs are append-only JSON lines, re-analysable offline.
Because \texttt{shared\_data} is a host bind mount, ingest and
inter-tier I/O traverse the Docker Desktop file-sharing layer,
which accounts for the occasional millisecond-scale contention
spikes in Table~\ref{tab:timing}.

\emph{Per-actor configuration.}  Table~\ref{tab:profiles}
summarises the fifteen actors: five clean baselines and ten
attackers spanning the eight profiles, with payload distribution,
inter-arrival ($i\!a\!t$), and resulting event rate per node.
Every profile additionally inserts a $U(0.5,1.5)$\,s idle gap on
15\% of iterations and a $U(1.0,5.0)$\,s startup delay.

\begin{table}[t]
\centering
\caption{Device profiles used in the testbed.  Buffer limit = 30\,B.}
\label{tab:profiles}
\renewcommand{\arraystretch}{1.1}
\setlength{\tabcolsep}{3pt}
\footnotesize
\begin{tabular}{@{}l l l l r@{}}
\toprule
\textbf{Nodes} & \textbf{Profile} & \textbf{Payload \& tool} & \textbf{$i\!a\!t$ (s)} & \textbf{evt/min} \\
\midrule
node-1..5-clean   & normal    & $[10,30]$\,B, \textsf{safe\_send}              & $U(0.1, 4.0)$   &   29 \\
node-6-overflow   & overflow  & $[35,60]$\,B \textsf{vulnerable\_send}   & $U(4.0, 10.0)$   &   9 \\
node-7-timespoof  & timespoof & \textsf{time\_spoof\_send}$^{*}$ & $U(2.0, 6.0)$   &   15 \\
node-8-spammer    & spam      & legal payloads                                  & $U(0.005,0.05)$ & 2200 \\
node-9-lowslow    & low-slow   & 10\% borderline 31--32\,B                       & $U(0.5, 3.0)$   &   34 \\
node-10-fuzzer    & fuzzer    & 60\% malformed content                & $U(0.5, 2.0)$   &   48 \\
node-11-pulsing   & pulsing   & 20-safe bursts every 20--30\,s                  & burst, 50\,ms   &   48 \\
node-12..15-mixed & mixed     & weighted send mix$^{\dagger}$    & $U(0.2, 3.5)$   &   32 \\
\bottomrule
\end{tabular}
\\[2pt]
{\footnotesize $^{*}$\,Timestamp offset $\Delta{=}{\pm}300/600$\,s.\quad $^{\dagger}$\,Send weights $(0.6,0.2,0.2)$ over \{safe, ovfl, spoof\}.}
\end{table}

The \emph{monitor} container runs the L1 canonicaliser (tailing the shared events log) and the L2 monitor (consuming canonical events and writing alerts); a separate \emph{correlator} container tails the alert stream and drives the MonPoly and RTLola subprocesses, writing incidents.

\emph{Protocol.}
An orchestration script brings the stack up, sleeps
$T_{\mathit{run}}{=}600$\,s for the primary run, tears down,
and runs metric-collection over the three log files.  Reported
metrics: alert and incident counts, per-property device-attributable
rate, and alert-to-incident latency.

\subsection{Aggregate workload}
\label{sec:eval:aggregate}

Table~\ref{tab:aggregate} summarises the 600\,s primary run. 
Of the 6872 raw events, 72 (1.0\%) are lost at the shared-log$\to$L1
ingest boundary (partial-line reads) \emph{before} canonicalisation;
L1 then processes the 6800 it receives and emits 6800 canonical events,
one per input, consistent with the totality of $\mu_1$.  L2 consumes
all 6800 (no L1$\to$L2 loss) and produces 6861 output records, of which 941
are \emph{security alerts} and 5920 are forwarded \textsf{safe\_tx} events
required by P3.3 and P3.5. L3 emits 137 incidents. A framed message
bus with a start-of-stream barrier would close the ingest gap, and we list
it as the top engineering follow-up.

\begin{table}[t]
\centering
\caption{Aggregate workload over a 600\,s 15-actor run.
``L2 sec.\ alerts'' excludes forwarded \textsf{safe\_tx}.}
\label{tab:aggregate}
\renewcommand{\arraystretch}{1.1}
\setlength{\tabcolsep}{6pt}
\footnotesize
\begin{tabular}{@{}l r @{\hspace{2.5em}} l r@{}}
\toprule
\textbf{Tier / metric} & \textbf{Count} & \textbf{L2 break-down} & \textbf{Count} \\
\midrule
Total raw events (L1)    & 6872 & L2 sec.\ alerts (total)   &  941 \\
L2  outputs (to L3)      & 6861 & \hphantom{X}\textsf{time\_anomaly} & 371 \\
\hphantom{X}of which forwarded \textsf{safe\_tx} & 5920 & \hphantom{X}\textsf{dos\_spam}    & 219 \\
L3 incidents             & 137 & \hphantom{X}\textsf{fuzzing}       & 258 \\
Total / attacker devices & 15 / 10 & \hphantom{X}\textsf{overflow}  & 93 \\
\bottomrule
\end{tabular}
\end{table}

\subsection{L3 incident detection}
\label{sec:eval:l3}

Table~\ref{tab:l3-detection} reports per-property L3 incident
counts on the same run, together with the unique attacker devices
each property identifies.  We report raw incident counts rather
than normalising by a fixed denominator because each MFOTL property
fires once per matching past-window, so a single attacker device
can emit several incidents in one run.

\begin{table}[t]
\centering
\caption{Per-property L3 incident emission, 600\,s 15-actor
primary run.  \emph{Inc.} is the total incidents fired;
\emph{Dev.} is the number of unique attacker devices that
triggered the property; \emph{Win.} is the temporal/window bound.}
\label{tab:l3-detection}
\renewcommand{\arraystretch}{1.1}
\setlength{\tabcolsep}{5pt}
\footnotesize
\begin{tabular}{@{}l l l c c c@{}}
\toprule
\textbf{ID} & \textbf{Incident type} & \textbf{Engine} & \textbf{Inc.} & \textbf{Dev.} & \textbf{Win.} \\
\midrule
P3.1 & \textsf{apt\_indicator}      & MonPoly & 41 & 4  & 60\,s     \\
P3.2 & \textsf{coordinated\_attack} & MonPoly & 19  & 6 & 30\,s \\
P3.3 & \textsf{escalation\_pattern} & MonPoly & 41 & 4  & 10--60\,s \\
P3.4 & \textsf{persistent\_threat}  & MonPoly &  6 & 6   & 1\,h      \\
P3.5 & \textsf{silent\_node\_anomaly} & RTLola & 30  & 5  & 5\,s$^{*}$ \\
\bottomrule
\end{tabular}
\\[2pt]
{\footnotesize $^{*}$\,Measured from the triggering overflow alert.}
\end{table}

P3.1 fires on the four mixed-APT nodes (\textsc{node-12}--\textsc{node-15}); P3.4 fires once for each of six attackers that accumulate at least five overflow alerts within its 1\,h window during the 600\,s run. \textbf{P3.5 fires 30 times} across five devices, once per overflow
that is followed by $\geq 5$\,s of silence; per-incident latency 4.99--5.01\,s by RTLola construction. The count is dominated by
\textsc{node-6} (20 of 30), whose bursty overflow profile produces many overflow-then-silence episodes; the number of \emph{unique} silent devices is five.

\paragraph{Incident accounting and P3.2 witnesses.}
Table~\ref{tab:incident-classification} sorts the 137 incidents
into: attributable-true, -mixed, -false-positive, and unattributed.  In this controlled labelled run, all 137 attributable incidents (P3.1--P3.5) correspond to attacker-labelled devices.  Each P3.2 incident has $\geq 3$ attacker contributors and no clean contributor, checked directly against the incident log. We report \emph{witness completeness ratio}
$\mathrm{WCR}=137/137 = 100\%$ as a first-class metric.

\begin{table}[t]
\centering
\caption{Four-bucket accounting of the 137 L3 incidents on the
600\,s primary run. \emph{Attr.\ true}: attributed to genuine
attacker device(s) only; \emph{Attr.\ mixed}: attacker and clean
devices both contribute; \emph{Attr.\ FP}: attributed to a clean
device (false positive); \emph{Unattr.}: no device attribution.}
\label{tab:incident-classification}
\renewcommand{\arraystretch}{1.1}
\setlength{\tabcolsep}{4pt}
\footnotesize
\begin{tabular}{@{}l c c c c c@{}}
\toprule
\textbf{Property} & \textbf{Attr.\ true} & \textbf{Attr.\ mixed} & \textbf{Attr.\ FP} & \textbf{Unattr.} & \textbf{Total} \\
\midrule
P3.1     & 41  & 0 & 0 & 0 & 41 \\
P3.2     & 19   & 0 & 0 & 0 & 19 \\
P3.3     & 41   & 0 & 0  & 0 & 41  \\
P3.4     &  6  & 0 & 0  & 0 &  6  \\
P3.5     &  30   & 0 & 0  & 0 &  30  \\
\midrule
\textbf{Total} & \textbf{137} & \textbf{0} & \textbf{0} & \textbf{0} & \textbf{137} \\
\bottomrule
\end{tabular}
\end{table}

\paragraph{End-to-end witness example.}
One P3.3 incident on \textsc{node-12-mixed} traces cleanly back to
L1: four \textsf{safe\_send}s, a back-jumped \textsf{time\_spoof\_send}
(a \textsf{time\_anomaly}/P2.3 alert), and a
\textsf{vulnerable\_send} over the buffer limit (a
\textsf{buffer\_overflow}/\textsf{overflow}) together fire
$\textsf{escalation\_pattern}(\textsc{node-12-mixed})$; the witness
is the six L2 alerts within 60\,s and the six raw L1 events behind them.

\subsection{Ablation: contribution of each layer}
\label{sec:eval:ablation}

Theorem~1 covers P3.2 (logically inexpressible below L3); P3.1/P3.3/P3.4
are placed at L3 by design.  Operationally on the 600\,s
primary run: the per-event L1 tier classifies
events but expresses no temporal pattern; \emph{L1+L2} yields 941 attack
alerts (371 time-anomaly, 219 dos-spam, 258 fuzzing, 93 overflow) that name
the device but no cross-device pattern; \emph{L1+L2+L3(MonPoly)}
adds 107 multi-event/multi-device incidents (P3.1--P3.4) that lie outside the lower tiers;

\emph{L1+L2+L3
(MonPoly \& RTLola)} adds a further 30 P3.5 silent-node anomalies
that no MonPoly-only configuration can detect within a bounded delay, since an event-triggered engine emits the verdict only when a later-timestamped input arrives. 

In the controlled labelled run, all 137 device-attributable incidents correspond to attacker-labelled devices.
Removing L3(MonPoly) loses 41 P3.1 + 19 P3.2 + 41 P3.3 + 6 P3.4; removing L3(RTLola) loses the 30 P3.5
detections. 

\subsection{Computational cost}
\label{sec:eval:cost}
Across the 118 incidents carrying a single triggering alert (all but
the 19 coordinated P3.2 incidents, which aggregate over a device set),
alert-to-incident latency is 1.71\,s
mean (min 0, max 5.0\,s); the 30 P3.5 incidents each carry RTLola's fixed
5\,s window by construction and dominate the mean, while every MonPoly
incident is resolved within that same 5\,s bound.

\begin{table}[t]
\centering
\caption{Per-event wall-clock cost at L1 and L2, measured inside
the monitor container during the 600\,s primary run.}
\label{tab:timing}
\renewcommand{\arraystretch}{1.1}
\setlength{\tabcolsep}{5pt}
\footnotesize
\begin{tabular}{@{}l c c c c c c@{}}
\toprule
\textbf{Tier} & \textbf{N events} & \textbf{mean ($\mu$s)} & \textbf{p50 ($\mu$s)} & \textbf{p95 ($\mu$s)} & \textbf{p99 ($\mu$s)} & \textbf{max ($\mu$s)} \\
\midrule
L1 & 6800 & 56.6  & 32.8  & 253.0  & 329.6 & 679.1 \\
L2 & 6800 & 710.6 & 604.8 & 1494.8 & 1860.2 & 15066.6 \\
\bottomrule
\end{tabular}
\end{table}

L1's $\sim$33\,$\mu$s median reflects JSON parsing, three Boolean checks,
and the content well-formedness scan; L2 is an order of magnitude higher
(605\,$\mu$s median) because each event triggers up to three deque updates
plus per-alert file writes. The L2 p99 of 1.86\,ms leaves ample headroom
under a sub-second budget; the 15.1\,ms max is a one-off shared-volume
contention spike. L1's $\sim$12$\times$ gap between mean (56.6\,$\mu$s) and
max (679\,$\mu$s) is dominated by occasional GIL releases and Python GC
pauses; p99 (330\,$\mu$s) is a tighter steady-state bound. Measurements were taken on the host described in Section~\ref{sec:eval:data}; absolute $\mu$s figures are hardware-dependent and should be read as orders of magnitude rather than portable constants.

\begin{figure}[t]
\centering
\begin{tikzpicture}
\begin{axis}[
  width=0.8\columnwidth, height=3.4cm,
  xlabel={Elapsed run time (s)},
  ylabel={Cum.\ incidents},
  xmin=0, xmax=620, ymin=0, ymax=45,
  grid=major, grid style={dashed,gray!30},
  legend pos=outer north east,
  legend style={font=\scriptsize, draw=none, fill=none, inner sep=2pt},
  tick label style={font=\scriptsize},
  label style={font=\scriptsize},
]
\addplot+[mark=*, mark size=1.2pt] coordinates {
 (0,0)(50,2)(100,5)(150,9)(200,13)(250,16)(300,21)(350,25)(400,25)(450,29)(500,33)(550,37)(600,41)};
\addlegendentry{P3.1}
\addplot+[mark=square*, mark size=1.2pt] coordinates {
 (0,0)(50,0)(100,2)(150,4)(200,6)(250,7)(300,9)(350,11)(400,12)(450,14)(500,16)(550,17)(600,19)};
\addlegendentry{P3.2}
\addplot+[mark=triangle*, mark size=1.4pt] coordinates {
 (0,0)(50,2)(100,5)(150,9)(200,13)(250,16)(300,21)(350,25)(400,25)(450,29)(500,33)(550,37)(600,41)};
\addlegendentry{P3.3}
\addplot+[mark=diamond*, mark size=1.4pt] coordinates {
 (0,0)(50,0)(100,0)(150,0)(200,1)(250,3)(300,4)(350,4)(400,5)(450,5)(500,5)(550,6)(600,6)};
\addlegendentry{P3.4}
\addplot+[mark=pentagon*, mark size=1.4pt] coordinates {
 (0,0)(50,1)(100,3)(150,5)(200,8)(250,11)(300,14)(350,16)(400,19)(450,22)(500,25)(550,27)(600,30)};
\addlegendentry{P3.5}
\end{axis}
\end{tikzpicture}
\caption{Cumulative L3 incidents vs.\ elapsed time (600\,s primary run, 50\,s
sampling); incidents accumulate roughly linearly as attackers persist.}
\label{fig:cumulative}
\end{figure}
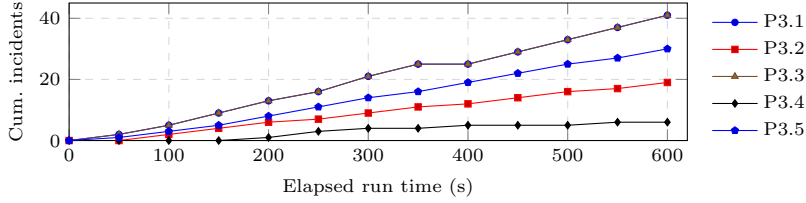

\emph{Accumulation over time.} Figure~\ref{fig:cumulative} plots
cumulative incidents per property over elapsed time. Rather than saturating,
all five properties accumulate steadily as the attackers persist through the
run, so absolute counts scale with run length (P3.4, on a 1\,h window, grows
most slowly); this is the run-length bound discussed next.

\emph{Bounding factors.}  Counts are bounded by run length
(short $T_{\mathit{run}}$ truncates the 1\,h P3.4 window) and by the
mixed-APT vector weights $(0.6,0.2,0.2)$, which make joint vector
pairs rare per 60\,s window; both raise P3.1 monotonically without
changing the spec.

\subsection{Coverage, evasion, and threats to validity}
\label{sec:eval:coverage}

\emph{Per-device coverage (false-negative analysis).}
P3.1 covers all four mixed-APT nodes; no gap is due to
mis-specification: every formula is satisfied iff the run is long
enough to accumulate the required evidence.  Specifically, P3.1 and
P3.3 each fire on all four mixed-APT nodes
(\textsc{node-12}--\textsc{node-15}, ${\sim}10\times$ each), P3.4 once
on each of six attackers, and P3.5 mostly on \textsc{node-6} (20 of
30); P3.2 draws on a contributing set of six (\textsc{node-6},
\textsc{node-9}, \textsc{node-12}--\textsc{node-15}), every incident
with $\geq 3$ contributors and none clean.
\emph{Evasion under known thresholds.}
An adversary knowing the L1/L2 thresholds has three evasion
options, each with a quantifiable cost.  \emph{(i)~Throttle
below a threshold:} to keep below $N_{\mathit{ovf}}{=}3$-in-30\,s
the attacker must drop to $\leq\!2$ overflow events per 30\,s,
a $\sim\!6\times$ slowdown over an unconstrained attacker (mean
inter-arrival $\approx 3$\,s in the \texttt{overflow} profile);
the analogous argument applies for spam vs.\ $N_{\mathit{spam}}{=}30$.
\emph{(ii)~Stay within a single L3 window:} single-vector attacks
on one device evade P3.1--P3.3 by construction, but the
throttling cost of (i) still applies at L1/L2.
\emph{(iii)~Coordinate across one device only:} P3.2 requires
overflow alerts on $\geq\!3$ distinct devices within 30\,s, so a
single-device adversary cannot trigger the predicate but also
cannot mount the multi-device threat the predicate targets.  The
framework therefore converts single-device evasion into a
constraint on attack \emph{breadth} ($\geq\!3$ devices) and
per-device \emph{slowdown} $\lceil
(\text{nominal-rate})/(\text{threshold-rate})\rceil$, supporting
the cost-asymmetry claim of Section~\ref{sec:intro}.

\label{sec:eval:iot23}
\emph{Real-data replay (IoT-23).}
We replay one capture of the IoT-23 dataset~\cite{iot23dataset}
(a Zeek log from a Mirai-infected device, ${\approx}23$\,k
connections) as a \mbox{compatibility/stress} test, not a
detection-accuracy validation: its per-connection Benign/Malicious
labels do not align with our event-level catalogue.  A coarse
adapter maps each connection to our schema (source IP as device,
IP-byte count as payload, Ethernet-MTU as buffer limit; failed
half-open connections trip the fuzz predicate; time-spoof disabled
for historical timestamps).  The L2 alert stream names a single
source IP, the capture's ground-truth-malicious device (source-IP
precision $1/1$); integrating a new dataset is a dataset-specific adapter
that emits the Section~\ref{sec:spec:schema} event schema, with L1/L2 unmodified.

\emph{Threats to validity.}
The dataset is synthetic and the eight attack classes are a closed
set; physical-hardware deployment is a follow-up
(Section~\ref{sec:discussion}).  Single-host containers compress
network jitter and clock skew present on a real multi-gateway
deployment.
\emph{Absence of a synchronisation layer.}
A Python supervisor over a shared log file dispatches events
between tiers; there is no dedicated synchronisation layer.
Four limitations follow: (i)~the shared-log dataflow is the
source of the L1$\to$L2 event-loss gap above; (ii)~one-second
timestamp resolution means two devices emitting in the same
second can reach the log out of true order, which MonPoly's
monotone-timestamp assumption tolerates only at low event
rates; (iii)~the RTLola subprocess runs on the system clock, so P3.5 deadlines fire even with no further input; however, events are timestamped on arrival rather than with
their upstream timestamps, so pipeline latency shifts the 5\,s window and
incident times are reconstructed as overflow-time\,+\,5\,s; (iv)~writes to the five
L3 subprocesses are sequential and unflow-controlled.  A proper
event-bus design (at-least-once delivery, per-consumer offsets,
periodic clock-tick) closes all four.
\emph{Threshold calibration, sensitivity, and benign bursts.}
The L1/L2 thresholds
($N_{\mathit{ovf}}{=}3$, $N_{\mathit{spam}}{=}30$,
$\tau_{ts}{=}15$\,s, $W_{\mathit{ovf}}{=}30$\,s,
$W_{\mathit{spam}}{=}5$\,s) are tuned to the testbed and not
claimed to generalise.  A quantitative sweep over
$N_{\mathit{ovf}}, N_{\mathit{spam}}, \tau_{ts}$, and $W_{L2}$,
reporting detection, FP, and latency at each operating point,
is queued as follow-up; Theorem~1 itself is
parameter-independent, but precision/recall numbers are not.
Real workloads also exhibit \emph{benign} bursts (firmware
updates, post-reconnect catch-up) whose statistics can resemble
attack patterns: our \texttt{pulsing} profile is benign-shaped
yet trips P2.2 at current thresholds, so deployment-time
calibration of $N_{\mathit{spam}}, N_{\mathit{pulse}},
W_{\mathit{spam}}$ against the actual workload is a prerequisite.

% ============================================================================
\section{Related Work}
\label{sec:related}

\emph{Hierarchical CPS-RV and runtime assurance.}
STREL~\cite{strel2017} and SaSTL~\cite{sastl2020} formalise
hierarchical spatio-temporal monitoring; we reuse the
decomposition but replace propositional spatial operators with
parametric MFOTL for cross-node correlation.  SOTER~\cite{desai2019soter}
and safe-RL shielding~\cite{shielding2018} pair monitors with
controller fallback.  Temperekidis
et al.~\cite{temperekidis2022fmirv} embed RV in FMI-based
co-simulation, a design-time analogue to our deployment-time
edge tier.

\emph{Comparison of RV engines for L3.}
The cloud tier needs two capabilities at once: cross-device
first-order quantification (to even write properties like ``three
different devices overflow within 30\,s'') and the ability to flag
a silent attacker when no event has arrived.  MonPoly~\cite{monpoly2014}
and the Isabelle-verified VeriMon~\cite{verimon2021} cover the first
natively but are event-triggered, detecting absence only when a later
event arrives, without a delay bound; RTLola~\cite{rtlola2019} is
time-triggered and handles the silent-node case but expresses
parametric joins less ergonomically.
TeSSLa~\cite{tessla2018,tessla2024}'s static parametrization scales
poorly to runtime-varying fleet sizes, and Copilot~\cite{pike2017copilot}
targets single-device embedded instrumentation.  Pairing MonPoly with
RTLola is the minimal combination covering both capabilities with
mature, online-capable engines.

\emph{Runtime verification in digital twins.}
A complementary line integrates RV with digital-twin (DT)
executions: Temperekidis et al.~\cite{temperekidis2022dt} sketch
a DT architecture with formal analysis for learning-enabled
autonomous systems~\cite{bensalem2024continuous,eleftheriadis2022nnequiv}, instantiated for traffic-sign
recognition~\cite{abdelsalam2024dttraffic} and depth-of-anesthesia
control~\cite{abdelsalam2025dtanesthesia}.  These share the
per-tier formal-spec design at \emph{design} time over a DT
trace; our framework moves the same discipline to \emph{deployment}
time over a fleet of edge devices.

\emph{IoT security monitoring and layered architectures.}
Cavalli et al.~\cite{cavalli2024iotsec} survey
\mbox{three-,} \mbox{four-,} and \mbox{five-layer} IoT
monitoring reference models, arguing no single-layer monitor
suffices.  We differ on three axes:
(i)~we give each tier a formal specification in a chosen logic
(boolean / past-MTL / MFOTL), making Theorem~1 provable from the
logical fragment of each tier; (ii)~we instantiate each attack
class on a 15-actor Docker testbed; (iii)~our composition tags
every L3 incident with the contributing L1/L2 sub-trace, giving
the audit record required by NIST SP~800-160 Vol.~2 and DORA.

\emph{IDS, resiliency frameworks, and audit alignment.}
Signature- and ML-classifier IDS~\cite{santhoshkumar2023mliot} and
uncertainty-aware classifier monitoring~\cite{lukina2021into} are
complementary to L1: a learned classifier could populate the
canonical-event stream without changing L2 or L3, though deploying
such CNN detectors on edge hardware carries its own performance
trade-offs~\cite{chintri2026edge}.  Cyber-resiliency
catalogues such as NIST SP~800-160
Vol.~2~\cite{nist800160v2} and NIST Zero-Trust~\cite{nist800207}
enumerate techniques but
stop short of numerical cost models; our framework instantiates
one such technique with measured costs.

% ============================================================================
\section{Discussion}
\label{sec:discussion}

\emph{Limitations.}  The 15-actor dataset is synthetic and
the eight attack classes are a closed set; real-world traffic
mixes are heavier-tailed and would require additional L1
predicates or threshold tuning.  Drift on the
underlying packet distribution may degrade the L1 thresholds over
time and is not yet modelled.  Sub-second event timestamps would
remove a class of near-coincident-event false positives observed
on coarse-grained testbed logs.

\emph{Partial compromise of L1.}
If an attacker bypasses the L1 canonicaliser on device $d$ and
emits raw events directly to the gateway log, L2's sliding counts
(P2.1, P2.2) and L3's properties (P3.1--P3.3) still apply, since
both consume the \emph{observed} stream, not the L1
verdict.  The residual risk is silent-bypass (dropping events
entirely); an L2 heartbeat predicate (``$d$ emitted
$\geq\!1$ event in the past 30\,s'') is a planned mitigation.

\emph{Artefact and reproducibility.}
The artefact bundles the three monitor programs, the four MonPoly
specs and the RTLola P3.5 spec, the device emulator, and the 15-actor
Docker Compose topology.  A single shell command rebuilds the image,
runs the stack, and emits JSON metrics and analysis files.

\emph{Future directions.}
(i)~Hardware deployment on Raspberry-Pi-class devices to
replicate L1 timing off the container host.
(ii)~Live-deployment, closed-loop evaluation measuring
throughput and operator workload under gateway-side APT alerts.
(iii)~Multi-gateway aggregation for fleets larger than one
gateway's capacity.
(iv)~Replacing the shared-log supervisor with an acknowledged event bus (at-least-once delivery, ordering barrier, periodic clock ticks), closing the
synchronisation limitations of Section~\ref{sec:eval}.
(v)~Automated property-catalogue synthesis from natural-language
procedural manuals (cf.\ \cite{mantra2026}).

% ============================================================================
\section{Conclusion}
\label{sec:conclusion}

We have presented a three-layer hierarchical security-monitoring
framework for edge-IoT, with a hybrid cloud tier that pairs
MonPoly with RTLola.  The pairing is what allows the framework
to close two gaps that show up together in practice and that no
single RV engine handles cleanly: cross-device quantification,
which becomes logically necessary once the gateway is
restricted to single-device state, and
absence-of-event detection, which becomes necessary as soon as
an attacker can compromise an edge monitor and go quiet.  On the controlled labelled 15-actor Docker testbed, every cloud incident the
framework attributes to a device names an attacker-labelled device, the
RTLola tier catches silent-bypass attempts that the MFOTL tier
cannot detect within a bounded delay, and per-tier monitoring cost stays well below typical
real-time budgets.

\subsubsection*{Acknowledgements.}
This work has received funding from the European Union's Digital Europe
Programme under grant agreement No~101190251 (CYBERGUARD).

% ---- Disclosure of Interests (Springer requirement; 9-point, 3rd-level head) -
{\small
\subsubsection*{Disclosure of Interests.}
The authors have no competing interests to declare that are relevant to the
content of this article.
}

% ============================================================================
\bibliographystyle{splncs04}
\bibliography{references}

\begin{thebibliography}{10}
\providecommand{\url}[1]{\texttt{#1}}
\providecommand{\urlprefix}{URL }
\providecommand{\doi}[1]{https://doi.org/#1}

\bibitem{abdelsalam2025dtanesthesia}
AbdElSalam, M., Bensalem, S., Delacourt, A., He, W., Katsaros, P., Kekatos, N.,
  Nolasco~Ruiz, R., Peled, D., Ponchant, M., Ryad, I., Temperekidis, A., Wu,
  C.: Digital twin for the formal analysis of a depth-of-anesthesia controller.
  SIMULATION  \textbf{{\mdseries 101}}(3),  341--360 (2025).
  \doi{10.1177/00375497241311617}

\bibitem{abdelsalam2024dttraffic}
AbdElSalam, M., et~al.: A digital twin prototype for traffic sign recognition
  of a learning-enabled autonomous vehicle. In: Companion Proc.\ of PoEM 2023.
  CEUR Workshop Proceedings, vol.~3645. CEUR-WS.org (2024).
  \doi{10.48550/arXiv.2402.09097}

\bibitem{shielding2018}
Alshiekh, M., Bloem, R., Ehlers, R., K{\"o}nighofer, B., Niekum, S., Topcu, U.:
  Safe reinforcement learning via shielding. In: AAAI Conf.\ on Artificial
  Intelligence (2018). \doi{10.1609/aaai.v32i1.11797}

\bibitem{mantra2026}
Anand, A., Chatzi, I., Raha, R., Schmuck, A.K.: {MANTRA}: Synthesizing
  {SMT}-validated compliance benchmarks for tool-using {LLM} agents. arXiv
  preprint arXiv:2605.06334  (2026). \doi{10.48550/arXiv.2605.06334}

\bibitem{strel2017}
Bartocci, E., Bortolussi, L., Loreti, M., Nenzi, L.: Monitoring mobile and
  spatially distributed cyber-physical systems. ACM-IEEE Intl.\ Conf.\ on
  Formal Methods and Models for System Design (MEMOCODE)  (2017).
  \doi{10.1145/3127041.3127050}

\bibitem{monpoly2014}
Basin, D., Klaedtke, F., M{\"u}ller, S., Z{\u{a}}linescu, E.: {MonPoly}:
  Monitoring usage-control policies. In: Runtime Verification (RV 2011). LNCS,
  vol.~7186. Springer (2012). \doi{10.1007/978-3-642-29860-8_27}

\bibitem{bensalem2024continuous}
Bensalem, S., Katsaros, P., Ni{\v{c}}kovi{\'c}, D., Liao, B.H.C., Ruiz~Nolasco,
  R., et~al.: Continuous engineering for trustworthy learning-enabled
  autonomous systems. In: Bridging the Gap Between AI and Reality (AISoLA
  2023). LNCS, vol. 14380, pp. 256--278. Springer (2024).
  \doi{10.1007/978-3-031-46002-9_15}

\bibitem{bonakdarpour2013ttrv}
Bonakdarpour, B., Navabpour, S., Fischmeister, S.: Time-triggered runtime
  verification. Formal Methods in System Design  \textbf{43}(1),  29--60
  (2013). \doi{10.1007/s10703-012-0182-0}

\bibitem{cavalli2024iotsec}
Cavalli, A.R., Mallouli, W., Montes~de Oca, E., Rivera, D.: {IoT} security
  monitoring tools and models. In: Ziegler, S., Rad{\'o}cz, R.,
  Quesada~Rodriguez, A., Matheu~Garcia, S.N. (eds.) Springer Handbook of
  Internet of Things, pp. 289--315. Springer Handbooks, Springer Nature
  Switzerland, Cham (2024). \doi{10.1007/978-3-031-39650-2_13}

\bibitem{chintri2026edge}
Chintri, M., Seitoglou, I., et~al.: Edge-oriented evaluation of {CNN}-based
  vehicle detection on embedded {V2X} platforms. In: Conf.\ on Innovation in
  Clouds, Internet and Networks (ICIN). pp.~1--5. IEEE, Athens, Greece (2026).
  \doi{10.1109/ICIN69025.2026.11481844}

\bibitem{tessla2018}
Convent, L., Hungar, S., Leucker, M., Scheffel, T., Schmitz, M., Thoma, D.:
  {TeSSLa}: Temporal stream-based specification language. In: Brazilian
  Symposium on Formal Methods (SBMF) (2018). \doi{10.1007/978-3-030-03044-5_10}

\bibitem{desai2019soter}
Desai, A., Ghosh, S., Seshia, S.A., Shankar, N., Tiwari, A.: {SOTER}: A runtime
  assurance framework for programming safe robotics systems. In: IEEE/IFIP
  Intl.\ Conf.\ on Dependable Systems and Networks (DSN) (2019).
  \doi{10.1109/DSN.2019.00027}

\bibitem{eleftheriadis2022nnequiv}
Eleftheriadis, C., Kekatos, N., Katsaros, P., Tripakis, S.: On neural network
  equivalence checking using {SMT} solvers. In: Formal Modeling and Analysis of
  Timed Systems (FORMATS) (2022). \doi{10.1007/978-3-031-15839-1_14}

\bibitem{falcone2021taxonomy}
Falcone, Y., Krsti{\'c}, S., Reger, G., Traytel, D.: A taxonomy for classifying
  runtime verification tools. International Journal on Software Tools for
  Technology Transfer (STTT)  \textbf{23}(2),  255--284 (2021).
  \doi{10.1007/s10009-021-00609-z}

\bibitem{rtlola2019}
Faymonville, P., Finkbeiner, B., Schledjewski, M., Schwenger, M., Stenger, M.,
  Tentrup, L., Torfah, H.: {StreamLAB}: Stream-based monitoring of
  cyber-physical systems. In: Computer Aided Verification (CAV) (2019).
  \doi{10.1007/978-3-030-25540-4_24}

\bibitem{iot23dataset}
Garcia, S., Parmisano, A., Erquiaga, M.J.: {IoT-23}: A labeled dataset of
  malicious and benign {IoT} network traffic (version 1.0.0).
  \url{https://www.stratosphereips.org/datasets-iot23} (2020), capture
  \texttt{CTU-IoT-Malware-Capture-34-1}

\bibitem{tessla2024}
Kallwies, H., Leucker, M., Schmitz, M., Schulz, M., Thoma, D., Weiss, A.:
  {TeSSLa}: An ecosystem for runtime verification. In: Runtime Verification (RV
  2022). LNCS, vol. 13498. Springer (2022). \doi{10.1007/978-3-031-17196-3_20}

\bibitem{leucker2009brief}
Leucker, M., Schallhart, C.: A brief account of runtime verification. Journal
  of Logic and Algebraic Programming  \textbf{78}(5),  293--303 (2009).
  \doi{10.1016/j.jlap.2008.08.004}

\bibitem{lukina2021into}
Lukina, A., Schilling, C., Henzinger, T.A.: Into the unknown: Active monitoring
  of neural networks. In: Runtime Verification (RV) (2021).
  \doi{10.1007/978-3-030-88494-9_3}

\bibitem{sastl2020}
Ma, M., Bartocci, E., Lifland, E., Stankovic, J.A., Feng, L.: {SaSTL}: Spatial
  aggregation signal temporal logic for runtime monitoring in smart cities. In:
  ACM/IEEE Intl.\ Conf.\ on Cyber-Physical Systems (ICCPS) (2020).
  \doi{10.1109/ICCPS48487.2020.00013}

\bibitem{pike2017copilot}
Pike, L., Goodloe, A., Morisset, R., Niller, S.: {Copilot}: A hard real-time
  runtime monitor. In: Runtime Verification (RV 2010). LNCS, vol.~6418.
  Springer (2010). \doi{10.1007/978-3-642-16612-9_26}

\bibitem{nist800207}
Rose, S., Borchert, O., Mitchell, S., Connelly, S.: Zero trust architecture.
  NIST Special Publication 800-207, National Institute of Standards and
  Technology (2020). \doi{10.6028/NIST.SP.800-207}

\bibitem{nist800160v2}
Ross, R., Pillitteri, V., Graubart, R., Bodeau, D., McQuaid, R.: Developing
  cyber-resilient systems: A systems security engineering approach. NIST
  Special Publication 800-160 Vol. 2 Rev. 1, National Institute of Standards
  and Technology (2021). \doi{10.6028/NIST.SP.800-160v2r1}

\bibitem{ukraine2015}
{SANS Industrial Control Systems and Electricity Information Sharing and
  Analysis Center (E-ISAC)}: Analysis of the cyber attack on the {Ukrainian}
  power grid: Defense use case. Tech. rep., SANS ICS / E-ISAC (Mar 2016)

\bibitem{santhoshkumar2023mliot}
Santhosh~Kumar, S.V.N., Selvi, M., Kannan, A.: A comprehensive survey on
  machine learning-based intrusion detection systems for secure communication
  in the internet of things. Computational Intelligence and Neuroscience
  \textbf{2023},  8981988 (2023). \doi{10.1155/2023/8981988}

\bibitem{verimon2021}
Schneider, J., Basin, D., Krsti{\'c}, S., Traytel, D.: {VeriMon}: Verifying
  {MFOTL} monitors. In: Computer Aided Verification (CAV). LNCS, Springer
  (2021)

\bibitem{temperekidis2022fmirv}
Temperekidis, A., Kekatos, N., Katsaros, P.: Runtime verification for
  {FMI}-based co-simulation. In: Runtime Verification (RV). LNCS, vol. 13498.
  Springer (2022)

\bibitem{temperekidis2022dt}
Temperekidis, A., Kekatos, N., Katsaros, P., et~al.: Towards a digital twin
  architecture with formal analysis capabilities for learning-enabled
  autonomous systems. In: MESAS 2022. LNCS, vol. 13866, pp. 163--181. Springer
  (2023). \doi{10.1007/978-3-031-31268-7_10}

\end{thebibliography}

\end{document}